\documentclass[12pt]{article}
\usepackage{epsf}
\newcommand{\beq}{\begin{equation}}
\newcommand{\eeq}{\end{equation}}
\newcommand{\beqa}{\begin{eqnarray}}
\newcommand{\eeqa}{\end{eqnarray}}

\newcommand{\no}{\nonumber}

\newcommand{\Mp}{M_{\pi^\pm}}
\newcommand{\Mn}{M_{\pi^0}}
\newcommand{\vs}{\vspace{-0.25cm}}

\begin{document}

\hfill {\small FZJ-IKP(TH)-1999-21} 



\bigskip\bigskip\bigskip

\begin{center}

{{\Large\bf Virtual photons in the pion form factors \\[0.3em]
   and the energy--momentum tensor}}

\end{center}

\vspace{.2in}

\begin{center}
{\large  Bastian  Kubis,\footnote{email: b.kubis@fz-juelich.de} 
Ulf-G. Mei{\ss}ner\footnote{email: Ulf-G.Meissner@fz-juelich.de}}

\bigskip

\bigskip

{\it Forschungszentrum J\"ulich, 
Institut f\"ur Kernphysik (Theorie)\\ 
D-52425 J\"ulich, Germany}

\end{center}

\vspace{.7in}

\thispagestyle{empty} 

\begin{abstract}
\noindent
We evaluate the vector and scalar form factor of the pion in the presence of virtual 
photons at next--to--leading order in two--flavor chiral perturbation theory. 
We also consider the scalar and tensor pion form factors 
of the energy--momentum tensor.
We find that the intrinsic electromagnetic corrections are 
of the expected size for the vector form factor 
and very small for the charged pion scalar form factor. 
Detector resolution independent photon corrections reduce the vector radius
by about one percent.
The scalar radius of the neutral pion is reduced by two percent by 
isospin--breaking contributions.
We perform infrared regularization by considering
electron--positron annihilation into pions and the decay of a light Higgs boson 
into a pion pair. We discuss the detector resolution dependent contributions to
the various form factors and pion radii. 
\end{abstract}

\vspace{0.8in}

\centerline{PACS: 12.39.Fe, 13.40.Gp, 13.40.Ks, 14.40.Aq}

\centerline{Keywords: 
{\it Pion form factors}, {\it chiral perturbation theory}, {\it energy--momentum tensor}}

\vfill

\newpage

\section{Introduction}
\label{sec:intro}
\def\theequation{\arabic{section}.\arabic{equation}}
\setcounter{equation}{0}
In their seminal paper of 1984, Gasser and Leutwyler~\cite{GLAnn} set up
the scheme to investigate the consequences of the broken chiral symmetry 
of QCD in a systematic fashion. This method is called chiral perturbation
theory (CHPT) and rests on the fact that at low energies, the (pseudo)Goldstone
bosons (the pions)   interact weakly and saturate most $n$--point functions
almost completely. The underlying power counting and an outline of the
method had already been established by Weinberg in 1979~\cite{Wein}.
In ref.~\cite{GLAnn}, a whole series of applications
was considered to one--loop accuracy (next--to--leading order), 
like e.g.\ the vector and scalar form factor of the
pion, elastic pion--pion scattering, pion radiative decay and so on.
By now, all of the observables
and processes considered in~\cite{GLAnn} have been worked out 
to two--loop accuracy,
either by direct Feynman graph calculations or by making use of dispersive
methods.\footnote{Note that using dispersive methods does in general not
allow to study the quark mass dependence of observables.} 
We mention here the pion mass and decay constant~\cite{Buergi}, the
pion form factors~\cite{GM,BT}, and $\pi\pi$ scattering~\cite{bern,orsay}.
In most cases these two--loop corrections are sizeably smaller than the
corresponding one--loop terms, as expected by the power counting underlying
CHPT. We remark that the improved two--loop representation of the vector
form factor of the pion has been used to significantly decrease the
hadronic uncertainty for the muon $(g-2)$ prediction~\cite{Jetc}. In addition,
there is a series of upcoming experiments to pin down precisely the
$\pi\pi$ S--wave scattering lengths (via $K_{l4}$ decays at BNL and
DA$\Phi$NE or pionium lifetime measurements at CERN). However, as first
pointed out in ref.~\cite{MMS} and further quantified in ref.~\cite{KU},
electromagnetic corrections to elastic $\pi\pi$--scattering in the threshold
region are of the same size as the hadronic two--loop contributions. At the
level of the presently achieved accuracy, it is therefore mandatory to
include virtual photons and work out the dual effects of isospin violation
due to the light quark mass difference and the electromagnetic interaction.
This machinery was essentially started in refs.~\cite{DHW,Bij,U} for the three
flavor case in connection with the violation of Dashen's theorem (see 
also ref.~\cite{EGPR}), which
is of importance for extracting the quark mass ratios from the masses
of the Goldstone bosons. 

\medskip\noindent Here, we will be concerned with  the vector
and scalar form factor of the pion as well as pionic matrix elements of
the energy--momentum tensor in the presence of virtual photons. 
The scalar form factor is of interest for
various reasons. Due to its quantum numbers, its phase is related to the
strong $\pi\pi$ interaction in the isospin zero S--wave and it also
exhibits the unitary cusp at the opening of the two--pion threshold.
Furthermore, it gives a direct measure of the QCD quark mass term in the pion.
In addition, the scalar form factor plays an essential
role in the dispersive analysis of the so--called pion--nucleon 
$\sigma$--term, which allows one to extract information about the strangeness
component in the nucleon wave function. In the pion--nucleon sector,
isoscalar quantities can be very sensitive to strong and/or electromagnetic
isospin breaking, see e.g.~\cite{W77,MS,FMS,MM}. It is therefore mandatory to
evaluate the effects of virtual photons on this scalar--isoscalar quantity.
Similarly, the hadronic two--loop representation of the vector form factor
is extremely precise and has even been used to extract the pion charge
radius from space-- and time--like data at low energies~\cite{BT}. Here,
the celebrated Ademollo--Gatto theorem~\cite{AG} lets one expect that 
symmetry breaking corrections are fairly small, but their precise 
magnitude has so far not been worked out. Another set of form factors is 
related to the pion matrix elements of the energy--momentum tensor~\cite{DL}, 
which play a role in the decay of a light Higgs into two pions~\cite{DGL,RW}
or heavy quarkonia transitions into the ground state with the emission of a 
pion pair. Of particular interest is the trace of the energy--momentum
tensor which allows one to study the interplay of conformal and chiral
symmetry~\cite{LS} or the hadron mass composition~\cite{SVZ}. Clearly, it is of
interest to learn about the modifications induced by virtual photons,
although we  expect these to be small effects.
 
\medskip\noindent  Of course, whenever one deals with virtual photons,
one encounters infrared (IR) singularities due to the vanishing photon mass.
These divergences can be cured by taking into account the effects of
radiation of soft photons which have an energy below a given detector
resolution. For that, one has to consider cross sections or decay widths. In our
case, we consider the squares of the corresponding form factors since that
is the way they appear in the pertinent cross sections or decay
rates. For the vector
form factor, we consider the process $e^+e^- \to \pi^+\pi^-$. The less
easily accessible scalar form factor together with the form factor of the
trace anomaly can be studied in the decay of the Higgs boson into two pions. 
We note that to the order we are working, we have to consider only the 
radiation of exactly one soft photon from any one of the final--state
pion legs. Radiation of two or more soft photons is of higher order in
the fine structure constant.

\medskip\noindent The paper is organized as follows. In sect.~\ref{sec:Lem}
we briefly discuss the strong and electromagnetic effective chiral Lagrangians
underlying our calculation. We work in $SU(2)$ and to fourth order in the
chiral expansion, counting the electric charge $e$ as a small parameter like
external momenta or pion mass insertions. In sect.~\ref{sec:pion} we give 
the results for the scalar and vector form factor of the pion at 
next--to--leading order including virtual photons and discuss the various
checks on our calculation. The infrared regularization is performed by means of
a finite photon mass. In sect.~\ref{sec:R} we briefly review the effective
Lagrangian for pions and external fields 
coupled to gravity and include virtual photons. The
pion matrix elements of the energy--momentum tensor are calculated in 
sect.~\ref{sec:EMT} to fourth order in the chiral expansion including virtual
photons and compared with previous results (obtained in the isospin limit). 
The problem of infrared regularization 
is taken up in sect.~\ref{sec:IR}. Our results are presented and 
discussed in  sect.~\ref{sec:res}. A short summary  is given in  sect.~\ref{sec:sum}.

\section{Effective Lagrangian with pions and virtual photons}
\label{sec:Lem}
\def\theequation{\arabic{section}.\arabic{equation}}
\setcounter{equation}{0}
In this section we discuss the strong and electromagnetic effective
Lagrangians underlying our calculations. The various pieces have already
been discussed in the literature, so we will be brief and just collect the
terms actually needed in what follows. The effective Lagrangian has the
energy expansion
\beq
{\cal L}_{\rm eff} = {\cal L}^{(2, {\rm str})} + {\cal L}^{(4, {\rm str})} 
+ {\cal L}^{(2, {\rm em})} + {\cal L}^{(4, {\rm em})} + \ldots~,
\eeq 
where the superscripts $2,4$ refer to the chiral dimension and ``str'' and ``em''
denote the strong and the electromagnetic terms, respectively. The ellipsis
stands for higher order terms not considered here. Our power counting is based
on the observation that $e^2/4\pi \simeq M^2_\pi /(4\pi F)^2 \simeq 1/100$, 
with
$M_\pi$ the pion mass and $F$ the weak pion decay constant. This means that
we consider the electric charge as a small parameter similar to the external
momenta and pion mass insertions. Consequently, any term of the form
$e^{2n} \, q^{2m} \, M_\pi^{2l}$ with $n+m+l=2$ is counted as fourth order.
In what follows, we denote such a term as ${\cal O}(q^4)$ without differentiating between
external momenta, pion mass insertions, or powers of the electric charge. 

\medskip
\noindent At leading order, the strong Lagrangian is nothing but the non--linear
$\sigma$--model coupled to external sources,
\beq
{\cal L}^{(2, {\rm str})} = \frac{F^2}{4} \langle u_\mu u^\mu 
+ \chi_+ \rangle~,
\eeq
where $\langle \ldots \rangle$ denotes the trace in flavor space, $U(x) = u^2(x)$
parametrizes the pion fields, $u_\mu = iu^\dagger D_\mu U u^\dagger$ 
is the conventional pion viel--bein and $D_\mu$ the chiral covariant 
derivative in the presence of external vector and axial--vector fields.
Furthermore, 
$\chi = 2B_0(s+ip)$ subsumes the external scalar and pseudoscalar sources, and 
$\chi_+ = u^\dagger \chi u^\dagger + u \chi^\dagger u$.
The external scalar source contains the quark mass matrix, $s(x) = 
{\rm diag}(m_u , m_d) + \ldots \,$, and $B_0 = |\langle 0| \bar q q |0\rangle|
/ F^2$ measures the strength of the spontaneous symmetry breaking.
We assume $B_0 \gg F$ (standard CHPT).
Throughout, we work in the so--called $\sigma$--model gauge, which simplifies the
calculations. It is given by
\beq 
U(x) = \sigma (x) \, {\bf 1}_2 + i\, \vec{\tau} \cdot \vec{\pi} (x) /F~, \quad
 \sigma (x) = \sqrt{1 - \pi^2 (x) /F^2} ~,
\eeq
with ${\bf 1}_2$ the unit matrix in flavor space.
The relevant terms of the strong Lagrangian at 
next--to--leading order read (we use the notation of ref.~\cite{BGS})
\beq\label{L4s}
{\cal L}^{(4,{\rm str})} = \frac{l_3}{16} \langle \chi_+ \rangle^2 + 
\frac{i\, l_4}{4} \langle u_\mu \chi_-^\mu\rangle + \frac{i\,l_6}{4}
\langle f_+^{\mu\nu} [u_\mu , u_\nu]\rangle - \frac{l_7}{16} \langle
\chi_-\rangle^2~,
\eeq
with $\chi_- = u^\dagger \chi u^\dagger - u \chi^\dagger u$, 
$\chi_-^\mu = u^\dagger D^\mu \chi u^\dagger -u D^\mu
\chi^\dagger u$, $f^+_{\mu\nu} = u F_{\mu\nu} u^\dagger + u^\dagger
F_{\mu\nu} u$, and $F_{\mu\nu} = \partial_\mu A_\nu - \partial_\nu
A_\mu$ is the conventional photon field strength tensor.
The low energy constants (LECs) $l_3$ and $l_4$ appear in the chiral expansion
of the pion mass and pion decay constant (and thus in the scalar form factor),
whereas $l_6$ features prominently in the pion charge radius as discussed below.
Of particular interest to us is the term $\sim l_7$ since it embodies the
strong isospin violation $\sim (m_d-m_u)$. After standard renormalization, the
$l_i$ turn into scale--dependent, renormalized LECs called $l_i^r (\lambda)$.
As first proposed in ref.~\cite{GLAnn}, we work instead with scale--independent
LECs $\bar{l}_i$ obtained from the $l_i^r (\lambda)$ via
\beq\label{barl}
 l_i^r (\lambda) = \frac{\beta_i}{32\pi^2} \,\biggl[\bar{l}_i + 
 \log\frac{\Mn^2}{\lambda^2} \biggr]~, 
\eeq 
with $\lambda$  the scale of dimensional regularization and $\beta_i$ 
the pertinent $\beta$--functions. We remark that we use the neutral pion
mass as the reference scale since the charged to neutral pion mass difference
is almost entirely of electromagnetic origin.
Note that $l_7$ does not get renormalized 
($\beta_7 =0)$ and in that case we simply set $\bar{l}_7 = 16\pi^2 \, l_7$. 
This procedure is adopted for all other counterterms with vanishing 
$\beta$--functions. The numerical values of the LECs will be given later.   

\medskip
\noindent The leading order electromagnetic Lagrangian reads~\cite{EGPR}
\beq
{\cal L}^{(2, {\rm em})} = -\frac{1}{4} F_{\mu\nu} F^{\mu\nu} - \frac{1}{2a}
(\partial \cdot A)^2 + C\, \langle QUQU^\dagger \rangle~,
\eeq
with $Q={\rm diag}(2/3,-1/3)$ the quark charge matrix
and $a$ the gauge fixing parameter. In what
follows, we work in the Feynman gauge $a=1$.\footnote{Here, we have already
set $Q_L =Q_R=Q$ since we are not concerned with the construction of the
effective Lagrangian.} In the  $\sigma$--model gauge for the pions, the last
term of this Lagrangian leads only to a term quadratic in pion fields. 
Consequently, the LEC $C$ can be calculated from the charged to neutral pion 
mass difference,
\beq
(\Mp^2 - \Mn^2)_{\rm em} = 
(\delta M^2_\pi)_{\rm em} = \frac{2e^2C}{F^2} = 2Ze^2F^2~,\quad Z=\frac{C}{F^4}~. 
\eeq
We have $Z=0.89$ if we use for $F$ its value in the chiral limit, $F_0 = 88\,$MeV.
At next--to--leading order,
the relevant terms from the  electromagnetic Lagrangian read~\cite{U,MMS,KU}
\beqa
{\cal L}^{(4, \, {\rm em})} &=& k_1 \, F^4 \,\langle
QUQU^\dagger\rangle^2 +
 k_2 \, F^2 \, \langle QUQU^\dagger \rangle \langle D_\mu UD^\mu
U^\dagger \rangle \no \\
&+& k_3 \,F^2 \, \bigl( \langle U^\dagger D_\mu U
Q\rangle^2 + \langle D_\mu U U^\dagger Q \rangle^2 \bigr) 
+  k_4 \, F^2 \, \langle U^\dagger D_\mu U Q\rangle 
\langle D_\mu U U^\dagger Q \rangle \no \\
&+& k_7 \, F^2 \, \langle QUQU^\dagger\rangle \langle \chi U^\dagger +
U \chi^\dagger \rangle +  k_8 \, F^2  \, \langle (U^\dagger \chi - \chi^\dagger U)
(U^\dagger QUQ -QU^\dagger QU) \rangle \no\\
&+&  k_9 \, F^4 \,\langle Q^2 \rangle \langle QUQU^\dagger\rangle
+  k_{10} \, F^2\,  \langle Q^2 \rangle \langle D_\mu U D^\mu
U^\dagger \rangle +  k_{11} \,F^2 \, \langle Q^2 \rangle \langle
\chi U^\dagger + U\chi^\dagger \rangle \no \\
&+&  k_{14} \, F^2 \, \langle (\chi U^\dagger + U\chi^\dagger ) \, Q +
(\chi^\dagger U + U^\dagger \chi) \, Q \rangle \langle Q \rangle~.
\eeqa
Note that the covariant derivative $D_\mu$ contains external vector sources
and the virtual photon fields, see e.g.\ refs.~\cite{U,MMS}.
The terms proportional to $k_1$ and $k_9$ are only needed for the mass
renormalization of the charged pions and the trace of the
energy--momentum tensor (as discussed below). 
For estimates of some of the LECs $k_i$ needed in connection with the violation of
Dashen's theorem, see refs.~\cite{BiPr,Mous}.
As shown in ref.~\cite{KU}, one needs additional divergent terms in the
one--loop generating functional to renormalize the photon propagator,
\beq\label{Gterms}
{\cal Z}^{({\rm em}, \gamma)}_{\rm 1-loop} = -\frac{1}{(4\pi)^2\epsilon} \int d^4 x
\biggl( -\frac{1}{6} \langle G_{\mu\nu}^R U
G^{L,\mu\nu} U^\dagger \rangle - \frac{1}{12} \langle G_{\mu\nu}^R 
G^{R,\mu\nu} + G^L_{\mu\nu}G^{L,\mu\nu}\rangle 
+ \frac{1}{6} \langle Q
\rangle^2 \, F_{\mu\nu}F^{\mu\nu} \biggr)~,
\eeq
(where $\epsilon=d-4$) in terms of left/right--handed gauge fields 
(for more precise definitions,
we refer to app.~A of ref.~\cite{KU}). In our case, we can
substitute $G_{\mu\nu}^{R,L}$ by $QF_{\mu\nu}$ and set $U(x) =1$. Thus,
eq.~(\ref{Gterms}) boils down to one term of the form
\beq\label{Fnew}
{\cal L}^{(4, {\rm em}, \gamma)} = -\frac{k_\gamma}{4}\, e^2 \, F_{\mu\nu}F^{\mu\nu}~,
\eeq
with the corresponding $\beta$--function being $\beta_\gamma =
2/3$. From that, one easily deduces the $Z$--factor of the photon,
\beq
Z_\gamma = 1 +  \biggl( \frac{e}{4\pi}\biggr)^2 \, \frac{1}{3} \, (\bar{k}_\gamma -
1 ) ~,
\eeq
in terms of the renormalized, scale--independent low energy constant $\bar{k}_\gamma$.

\section{Pion form factors}
\label{sec:pion}
\def\theequation{\arabic{section}.\arabic{equation}}
\setcounter{equation}{0}
The pion vertex functions with one external vector or scalar source are parametrized
in terms of the vector and scalar form factor of the pion, respectively. The
vector form factor is experimentally accessible in electron scattering off pions
or electron-positron annihilation into a pair of pions. Due to the quantum numbers,
the first non--Goldstone meson, the $\rho (770)$, features prominently in this channel
and limits the range of applicability of CHPT. However, even at much lower energies,
the tail of the $\rho$ is visible, it saturates the LEC $\bar{l}_6$ and thus
largely determines the pion charge radius, as discussed below. The scalar form
factor can only be inferred indirectly due to the absence of any external scalar
isoscalar source. It plays a central role in the discussion of low energy theorems
since the scalar radius is intimately related to the change of the pion decay
constant when the quark masses are turned on, as pointed out in ref.~\cite{GLAnn}.
We now turn to a more detailed discussion of these two form factors.

\subsection{Vector form factor}
The vector form factor of the pion is defined by
\beq
\langle \pi^a (p') \pi^b(p) \, {\rm out} |V_\mu^c|0 \rangle = i\, \epsilon^{abc} \,
(p_\mu' - p_\mu) \, F_\pi^V (s)~,
\eeq
with $s = (p'+p)^2$ and $V_\mu^c$ the $c$--component of the quark
isovector vector current, $V_\mu^c = \bar{q} \, \gamma_\mu \, (\tau^c/2)\, q$. 
Gauge invariance requires $F_\pi^V (0)=1$.
Since the neutral pion is its own antiparticle, the corresponding form factor
vanishes identically. At next--to--leading order in the chiral
expansion, there is no strong isospin breaking in $F_\pi^V$~\cite{AG}. 
The low energy expansion of the vector form factor defines the
electromagnetic charge radius of the pion,
\beq
F_\pi^V (s) = 1 +\frac{1}{6} \, \langle r^2\rangle_\pi^V \, s + {\cal
  O}(s^2)~.
\eeq
The one--loop representation of $F_\pi^V$ has been given in~\cite{GLAnn}, whereas
the expansion to two loops was worked out using dispersive methods in~\cite{GM}
and by direct diagrammatic calculation in~\cite{BT}. Here, we are concerned
with the virtual photon effects at next--to--leading order (one--loop),
including in particular virtual photon loops.

\begin{figure}[htb]
\centerline{
\epsfysize=7.3cm
\epsffile{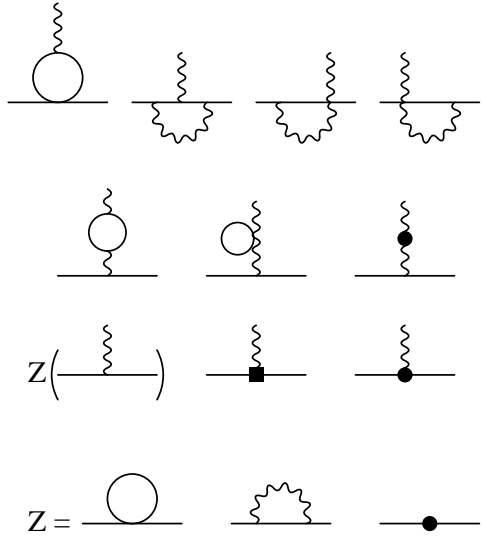}
}
\vskip 0.5cm
\caption{Vector form factor of the pion to one loop. Solid and wiggly lines denote
 pions and photons, in order. Strong and electromagnetic counterterms
 are depicted by the filled box and circle, respectively. Diagrams
 that vanish in dimensional regularization are not shown.
\label{FV}
}
\end{figure}
\medskip \noindent
The pertinent Feynman diagrams for a complete calculation at fourth order
with virtual photons are depicted in fig.~\ref{FV}. In the first row, we show
the strong one--loop graph together with the irreducible one--photon loop diagrams.
The second row contains the reducible photon loop diagrams, below
the standard wave function renormalization as well as strong and electromagnetic
counterterms are depicted. The graphical representation of the $Z$--factor
is also shown. We give here the $Z$--factors for the neutral and charged
pions (note that $Z_{\pi^0}$ is needed for the calculation of the
scalar form factor in the next paragraph),
\beqa\label{Zpi0}
Z_{\pi^0} &=& 1 - \frac{\Delta_{\pi^0} }{F^2} + e^2 \,
\biggl( 2 (2k_3+k_4) - \frac{20}{9} (k_2+k_{10} ) \biggr)~, \\\label{Zpip}
Z_{\pi^\pm} &=& 1 - \frac{\Delta_{\pi^\pm} }{F^2} +2 e^2 \, 
\frac{\Delta_{\pi^\pm} }{\Mp^2} - 2\biggl( \frac{e}{4\pi}\biggr)^2 \,
\biggl(1+  2 \log\frac{m_\gamma}{\Mp} \biggr) 
- \frac{20}{9} e^2 (k_2+k_{10})~, 
\eeqa
with
\beq
\Delta_{\pi^a} = 2\,M^2_{\pi^a} \, \biggl( L + \frac{1}{(4\pi)^2} \log
\frac{M_{\pi^a}}{\lambda} \biggr)~.
\eeq
Here, $L$ contains
the poles in $1/(d-4)$ as $d\to 4$. 
Note that the $k_i$ in eqs.~(\ref{Zpi0}, \ref{Zpip})
are the non-renormalized LECs of the dimension four counterterms.
The photon loop graphs lead to a complication, namely they are not
infrared finite. To handle these IR divergences related to  the zero photon mass, we
introduce a small photon mass $m_\gamma$, giving a modified photon propagator
\beq
D_{\mu\nu} (x) = \int \frac{d^dk}{(2\pi)^d} 
\frac{- i g_{\mu\nu}}{k^2 - m_\gamma^2 + i\epsilon}\,e^{-ik\cdot x}~.
\eeq
Observables are, of course, independent of $m_\gamma$, as discussed in 
section~\ref{sec:IR}. At fourth order, the vector form factor of the pion takes the form:
\beqa\label{FV1loop}
F_\pi^V(s) &=& 1 + \frac{1}{(4\pi F)^2}\, \frac{1}{6} 
\biggl\{\biggl(\bar{l}_6 - \frac{1}{3}\biggr)\, s 
+  (4\Mp^2-s) \, K_\pm (s) \biggr\} \no\\ 
&& \,\,\,\,  +\biggl( \frac{e}{4\pi}\biggr)^2 \, 
\biggr\{\frac{1}{3} \, \frac{28\Mp^2-13 s}{4\Mp^2 -s} \, K_\pm (s) 
- \frac{4s}{4\Mp^2-s}
- \frac{4}{3} \biggl( \frac{\Mp^2}{s} \, K_\pm (s) +
  \frac{1}{6} \biggr) \no \\
&& \qquad\qquad
+\frac{s-2\Mp^2}{\Mp^2} \, G(s)
+ 4 \, \biggl( \frac{s-2\Mp^2}{s\, \sigma}\,
\log \frac{\sigma+1}{\sigma-1} -1 \biggr) \, \log \frac{m_\gamma}{\Mp}
\biggr\}~,
\eeqa
with
\beq\label{sig}
\sigma = \sqrt{1 - \frac{4\Mp^2}{s}}
\eeq
and
\beqa\label{lfct1}
K_a (s) &=& \int_0^1 dx \log \biggl(1-x(1-x)\frac{s}{M^2_a} \biggr) ~,
\quad a=\{0,\pm\}~,\\ \label{lfct2}
G(s) &=& \int_0^1 dx \,
\frac{\log \Bigl(1-x(1-x)\frac{s}{\Mp^2}\Bigr)}{1-x(1-x)\frac{s}{\Mp^2}}~.
\eeqa
We note that the irreducible photon loop graphs are
singular at $s=4\Mp^2$ whereas the reducible photon loops are smoothly
varying with energy.  A representation of $G(s)$ in terms of
dilog--functions is given in
ref.~\cite{KU} (since $G(s)$ is related to the IR--finite part of
their loop function $G_{+-\gamma} (s)$).
The non--renormalization theorem of the electric charge is fulfilled since
all the various contributions are of ${\cal O}(s)$. This is easily
verified using
\beq
K_a (s) = -\frac{s}{6M_{\pi^a}^2} + {\cal O}(s^2)~, \quad
G(s) = -\frac{s}{6\Mp^2} + {\cal O}(s^2)~.
\eeq
In particular, the electromagnetic counterterms from the fourth order Lagrangian,
which are momentum--independent, cancel exactly with the corresponding counterterms
in the $Z$--factor, compare fig.~\ref{FV}. Concerning the strong contribution, it is
well--known that the counterterm $\sim \bar{l}_6$ dominates
the pion charge radius. This LEC is entirely saturated by the $\rho$--meson,
which means that even in the theory of pions only, this resonance plays a visible role.

\subsection{Scalar form factor}
The scalar form factor of the pion is defined by
\beq
\langle \pi^a (p') \pi^b (p) \, {\rm out} | m_u \bar{u}u + m_d \bar{d}d |0\rangle 
= \delta^{ab} \,  \tilde{\Gamma}_\pi (s)~.
\eeq
In order to display the momentum dependence, one introduces  the normalized 
scalar form factor
$\Gamma_\pi (s) =  \tilde{\Gamma}_\pi (s)/  \tilde{\Gamma}_\pi (0)$ so that
$\Gamma_\pi (0) = 1$. The low energy expansion of the normalized scalar form
factor defines the scalar radius of the pion,
\beq
\Gamma_\pi (s) = 1 +\frac{1}{6} \, \langle r^2\rangle_\pi^S\,s + {\cal
  O}(s^2)~.
\eeq
Note that while the electromagnetic radius is directly measurable in
electron scattering, the absence of scalar--isoscalar external sources
allows only for an indirect determination of the scalar radius by e.g.
a dispersive analysis of elastic $\pi\pi$ scattering~\cite{DGL}. 
At $s=0$, the scalar form factor is proportional to the pion expectation value
of the QCD quark mass term
\beqa
{\cal H}_{\rm QCD} = m_u \bar{u}u + m_d \bar{d}d &=& \frac{1}{2}(m_u+m_d)
(\bar{u}u + \bar{d}d ) + \frac{1}{2}(m_u-m_d) (\bar{u}u - \bar{d}d )~\no\\
&=&\quad\quad \hat{m} \, (\bar{u}u + \bar{d}d ) \qquad \,\,
+ \quad\quad \tilde m \, (\bar{u}u - \bar{d}d )~,
\eeqa
in terms of the isoscalar and isovector components.
Therefore one can apply the Feynman--Hellman theorem~\cite{RF,He} separately to
the isoscalar term,
\beq\label{FH1}
\frac{\partial M_\pi^2}{\partial \hat{m}} = \langle \pi | \bar{q}q | \pi \rangle~,
\eeq
and similarly to the isovector one,
\beq\label{FH2}
\frac{\partial M_\pi^2}{\partial \tilde{m}} = \langle \pi | \bar{q} \tau^3 q 
| \pi \rangle~.
\eeq
In the case of virtual photons, the arguments leading to
eqs.~(\ref{FH1}, \ref{FH2}) are not altered. These equations thus
serve as a powerful check on our calculations.
The one--loop representation of $\tilde{\Gamma}_\pi$ has been given 
in~\cite{GLAnn}, whereas the
expansion to two loops was worked out using dispersive methods in~\cite{GM}
and by direct diagrammatic calculation in~\cite{BT}. We are concerned
with the virtual photon effects at next--to--leading order. 
To the order we are working, strong isospin
violation only occurs in $\tilde{\Gamma}_\pi (0)$.

\medskip

\begin{figure}[htb]
\centerline{
\epsfysize=3.4cm
\epsffile{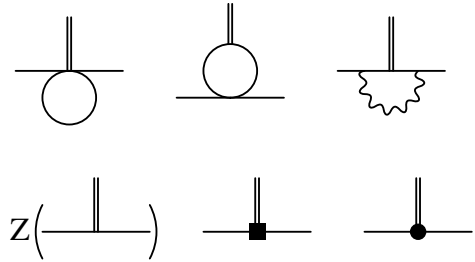}
}
\vskip 0.5cm
\caption{Scalar form factor of the pion to one loop. The double line denotes the
coupling to the scalar--isoscalar source. For further notations, see
  fig.~\ref{FV}.\label{SFF}}
\end{figure}
\noindent The pertinent tree level and one--loop Feynman diagrams needed to 
calculate the scalar form factor at ${\cal O}(q^4)$ are shown in fig.~\ref{SFF}.
The scalar form factor for the charged pions is given by
\beqa \label{GSch1loop}
\Gamma_{\pi^\pm}(s) &=& 
1 + \frac{1}{(4\pi F)^2} \biggl\{ s\, \biggl(\bar{l}_4 -1 -
\log\frac{\Mp}{\Mn} \biggr) + \frac{1}{2} (\Mn^2 -s)\, K_0(s) -\frac{s}{2}\,K_\pm (s)   
\biggr\} \no \\
&&  \qquad + \biggl( \frac{e}{4\pi}\biggr)^2 \, \biggl\{ (1-4Z)\,K_\pm (s) +
\frac{s-2\Mp^2}{\Mp^2}\, G(s) \no\\
&& \qquad \qquad\qquad + 4\biggl(\frac{s-2\Mp^2}{s \, \sigma}\,
 \log  \frac{\sigma+1}{\sigma-1} -1 \biggl) \,  
\log \frac{m_\gamma}{\Mp}\,\biggr\}~,\\
\tilde{\Gamma}_{\pi^\pm} (0) &=& \Mp^2 + \frac{\Mn^4}{(4\pi F)^2} \,
\frac{1}{2}(1-\bar{l}_3) - \biggl( \frac{e}{4\pi}\biggr)^2 \, \Mn^2 \,(3+4Z) \no\\
&& \qquad - 2Ze^2F^2 \biggl\{ 1 + \biggl( \frac{e}{4\pi}\biggr)^2
\biggl[ 4- (6+8Z) \log \frac{\Mp}{\Mn} + \frac{\bar{k}_\pm '}{Z} \biggr]\biggr\}~,
\eeqa
and similarly for the scalar form factor of the neutral pions
\beqa
\Gamma_{\pi^0}(s) &=& 
1 + \frac{1}{(4\pi F)^2} \biggl\{ s\, \biggl(\bar{l}_4 -1 -
2\log\frac{\Mp}{\Mn} \biggr) +  (\Mn^2 -s)\, K_\pm (s) -\frac{\Mn^2}{2}\,K_0 (s)   
\biggr\}~, \no \\ && \\
\tilde{\Gamma}_{\pi^0} (0) &=& \Mn^2 \biggl\{ 1+ \frac{\Mn^2}{(4\pi F)^2} \,
\biggl( \frac{1}{2}(1-\bar{l}_3) - 2\bar{l}_7 \, \delta^2  +
2\, \log \frac{\Mp}{\Mn} \biggr) \biggr\}~,
\eeqa
with
\beqa
\bar{k}_\pm ' &=& \frac{5}{6} \Bigl(1+2Z+8Z^2\Bigr) \, \bar{k}_1 
-\frac{20}{9}Z^2\bar{k}_2 - \frac{1}{6} \Bigl(5+Z+4Z^2\Bigr) \, \bar{k}_9 +
\frac{1}{18}\,\Bigl( 27+4Z \Bigr) Z
\,\bar{k}_{10}~, \\
\delta &=& \frac{m_u-m_d}{m_u+m_d} \approx -0.3 ~,
\eeqa
and the loop functions $K_a (s)$, $G(s)$ as given in the preceding paragraph.
The light quark mass difference only appears in the normalization of the
scalar form factor. This is related to the fact that isospin breaking only appears
at next--to--leading order, see the last term in eq.~(\ref{L4s}). This operator
only has non--vanishing matrix--elements with two neutral pions (similar to the
operators $c_5$ and $f_2$ which appear in neutral pion scattering off 
nucleons~\cite{MS,MM}). To arrive at the formulae for
$\tilde{\Gamma}_{\pi^0,\pi^\pm}$, we have made use of the
electromagnetic renormalization of the pion masses as given in refs.~\cite{MMS,KU}.

\section{Effective Lagrangian coupled to gravity}
\label{sec:R}
\def\theequation{\arabic{section}.\arabic{equation}}
\setcounter{equation}{0}

The energy--momentum tensor for chiral effective theories has been
studied in ref.~\cite{DL}. 
This extended Lagrangian can be constructed by including 
the metric tensor as an external source
which couples to the energy--momentum tensor. At next--to--leading order, one
has three new terms (plus additional terms without pions, needed for the
renormalization). The energy--momentum tensor plays a role in the pionic
decay channels of a hypothetical light Higgs particle via the matrix element
$\langle \pi\pi | \Theta_\mu^\mu | 0\rangle$. We will later use this
process to perform the IR regularization of the scalar pion form factor. 
Here, we briefly collect the pertinent results from ref.~\cite{DL}, 
adapted to our notation. In addition, we also include virtual photons.

\medskip
\noindent
We work in $SU(2)$ and rewrite the terms given in ref.~\cite{DL} in terms
of LECs $l_{11,12,13}$ (note that we keep the numbering as in~\cite{DL}).
The pertinent strong  Lagrangian coupled to gravity reads 
(terms just needed for  renormalization are not shown)\footnote{At leading
order one has no new terms. One simply raises and lowers the Lorentz indices 
of the non--linear $\sigma$--model with the metric tensor $g_{\mu\nu}$.}
\beq
{\cal L}^{(4, {\rm str, R})} = l_{11} \, R \, \langle D_\mu U D^\mu U^\dagger \rangle
+ l_{12} \, R^{\mu\nu} \, \langle D_\mu U D_\nu U^\dagger \rangle
+ l_{13} \, R \, \langle \chi U^\dagger + U\chi^\dagger \rangle~,
\eeq
where $R_{\mu\nu}$ and $R$ are the Ricci tensor and curvature scalar, respectively,
\beq
R_{\mu\nu} = R^\lambda_{\mu\lambda\nu}~, \quad R = R_{\mu}^{\mu}~,
\eeq
derived from the curvature tensor and the Christoffel symbols,
\beqa
R_{\sigma\mu\nu}^\lambda &=& \partial_\mu \Gamma_{\nu\sigma}^\lambda -
\partial_\nu \Gamma_{\mu\sigma}^\lambda + \Gamma_{\mu\alpha}^\lambda
\Gamma^\alpha_{\nu\sigma} - \Gamma^\lambda_{\nu\alpha}\Gamma^\alpha_{\mu\sigma}~,\\
\Gamma^\rho_{\mu\nu} &=& \frac{1}{2} \,g^{\rho\sigma} \, 
(\partial_\mu g_{\nu\sigma}
+ \partial_\nu g_{\mu\sigma} - \partial_\sigma g_{\mu\nu} )~.
\eeqa
{}From these
formulae it becomes obvious why $R$ and $R_{\mu\nu}$ count as ${\cal O}(q^2)$.
For the case of $SU(2)$, the $\beta$--functions for $l_{11}$ and
$l_{13}$ are 
\beq
\beta_{11} = \frac{1}{6}~, \quad \beta_{13} = \frac{1}{8}~.
\eeq 
Note that the coupling $l_{12}$ is not renormalized. 
For an estimate of the LECs and a more detailed discussion, see
ref.~\cite{DL}. As before, we introduce
scale--independent LECs $\bar{l}_{11,12,13}$ as described in
sect.~\ref{sec:Lem}. As numerical values, we use
\beq
\bar{l}_{11} = 8.7\, , \quad \bar{l}_{12} = -0.4\, , \quad 
\bar{l}_{13} = 9.4\, .
\eeq
These values have been obtained from the ones given in ref.~\cite{DL}
by running them down to $\lambda = \Mn$ using the $SU(3)$
$\beta$--functions and converting the obtained values for the
renormalized, scale--dependent LECs into the scale--independent ones
according to eq.~(\ref{barl}).

\medskip \noindent 
We are interested in the virtual photon corrections to the
pionic matrix elements of the energy--momentum tensor. There is exactly
one novel electromagnetic counterterm at fourth order,
\beq
{\cal L}^{(4, {\rm em,  R})} = k_{15} \, F^2 \, R \, \langle QUQU^\dagger \rangle~.
\eeq
Performing standard renormalization, the $\beta$--function related to the LEC
$k_{15}$ is
\beq
\beta_{15} = \frac{1}{2} + \frac{2}{3}\, Z~.
\eeq
It is obvious from the power counting that there is only one such term. 
The electric charge has to come in squared, and since $R$ is
of order $q^2$, we cannot have derivatives on the pion fields or quark
mass insertions.

\section{Pion form factors of the energy--momentum tensor}
\label{sec:EMT}
\def\theequation{\arabic{section}.\arabic{equation}}
\setcounter{equation}{0}
Consider now matrix elements of the energy--momentum tensor in one--particle
states like e.g.\ the pion. The general form of such matrix elements follows
from Lorentz invariance and four--momentum conservation~\cite{DL},
\beq
\langle \pi^a (p') | \Theta_{\mu\nu} | \pi^b(p)\rangle = \delta^{ab} \,
\biggl[\frac{1}{2} (s\,g_{\mu\nu} - q_\mu q_\nu )\, \Theta_1 (s)
+\frac{1}{2} P_\mu P_\nu \, \Theta_2 (s)\biggr]~,
\eeq
with $P_\mu =p_\mu'+p_\mu$, $q_\mu = p_\mu'-p_\mu$, and $s=q^2$. 
The two form factors reflect
the fact that the energy--momentum tensor contains a scalar and a tensor part.
In what follows, we are mostly interested in the matrix element of the trace
of the energy--momentum tensor, $\Theta_\mu^\mu$, which is a scalar. The
corresponding matrix element takes the form (neglecting isospin indices)
\beq\label{trace}
\langle \pi | \Theta_\mu^\mu | \pi \rangle \equiv \Theta_\pi (s)
= \frac{3}{2} s \, \Theta_1 (s) + \frac{1}{2}(4M_\pi^2-s)\, \Theta_2 (s)~.
\eeq
Because $\Theta_{00}$ is the Hamiltonian density, we require the normalization
$\Theta_2 (0) =1$. 

\medskip 

\begin{figure}[htb]
\centerline{
\epsfysize=5.9cm
\epsffile{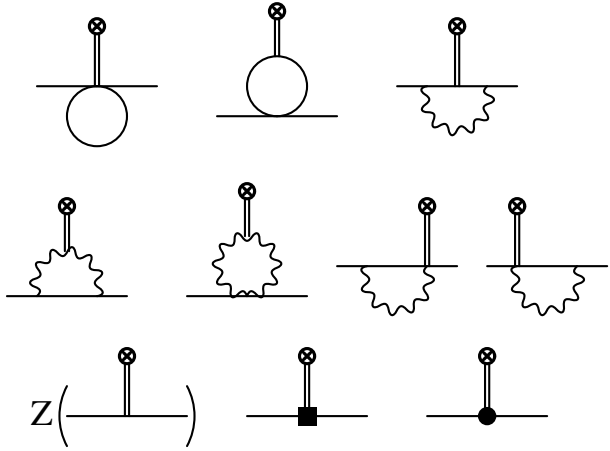}
}
\vskip 0.5cm
\caption{Pion form factors of the energy--momentum tensor to one loop. 
The double line with the circle-cross denotes the coupling to 
the energy--momentum tensor.  For further notations, see  fig.~\ref{FV}.\label{TRV}
}
\end{figure}
\noindent
The pertinent Feynman diagrams needed to evaluate the form factors of the 
energy--momen\-tum
tensor at next--to--leading order (including virtual photons) in one--pion states 
are shown in fig.~\ref{TRV}.  
At tree level, one simply has $\Theta_1 (s) = \Theta_2 (s)
= 1$. The full one--loop result for the charged pions reads
\beqa\label{t12c}
\frac{\Theta_1^{\pi^\pm} (s)}{\Theta_1^{\pi^\pm} (0)^{\rm fin}} &=& 1 -
\frac{1}{(4\pi F)^2} \, \biggl\{ \frac{8}{9} s- 2s \log\frac{\Mp}{\Mn}
+\frac{1}{3}(s+\Mn^2)\, K_0(s) +\frac{1}{3}(s+2\Mp^2)\, K_\pm (s) 
\no \\
&& \qquad\qquad\qquad -\frac{2}{3}\Mn^2 \biggl(\frac{\Mn^2}{s}K_0 (s)
+ \frac{1}{6}\biggr) -2s \biggl( \frac{1}{3}\bar{l}_{11} +
\bar{l}_{12} \biggr) \biggr\} \no\\
&& + \biggl(\frac{e}{4\pi}\biggr)^2 \, \biggl\{
\frac{8}{3}(1+Z)K_\pm (s) +  \frac{8}{3}(2Z-1) \biggl(\frac{\Mp^2}{s}K_\pm
(s)+ \frac{1}{6}\biggr) \no\\
&& \qquad\qquad\quad + \frac{s-2\Mp^2}{\Mp^2} \, G(s) + 
2\Mp^2 \,\Bigl(H(s) -2H_1(s)\Bigr)  \\ 
&&\qquad\qquad\quad- \frac{2}{3} \biggl(
\log \frac{s}{\Mp^2} -i\pi \biggr) 
 + 4\, \biggl( \frac{s-2\Mp^2}{s \,\sigma} \log 
\frac{\sigma+1}{\sigma-1} -1 \biggr) \, \log\frac{m_\gamma}{\Mp} \, 
\biggr\}~,\no \\
\Theta_1^{\pi^\pm} (0)^{\rm fin} &=& 1 + \frac{\Mn^2}{(4\pi F)^2} \,
\biggl\{ \frac{1}{3} -\frac{4}{3}\log\frac{\Mp}{\Mn} +\bar{l}_{13} - 
\frac{4}{3}\bar{l}_{11} \biggl\} \no \\
&&
+ \biggl(\frac{e}{4\pi}\biggr)^2 \,\biggl\{ \frac{8}{9}
(2+3Z) + \biggl(2 + \frac{8}{3}Z \biggr) \, 
\biggl(\bar{k}_{15} + 2\log\frac{\Mp}{\Mn} \biggr) \biggr\} ~,\\ 
\Theta_2^{\pi^\pm} (s) &=& 1 - \frac{2s}{(4\pi F)^2} \, \bar{l}_{12}
\no\\
&& + \biggl(\frac{e}{4\pi}\biggr)^2 \, \biggl\{
\frac{4s}{s-4\Mp^2} + 6 \frac{s-2\Mp^2}{s-4\Mp^2}\,K_\pm (s)
+ \frac{s-2\Mp^2}{\Mp^2} \, G(s) \no\\
&& \qquad\qquad\quad + \biggl(2\Mp^2 \,\Bigl(H(s) -2H_1(s)\Bigr) - \frac{2}{3} \biggl(
\log \frac{s}{\Mp^2} -i\pi \biggr) \biggr) \, \frac{3s}{s-4\Mp^2}
\no\\
&& \qquad\qquad\quad + 4\, \biggl( \frac{s-2\Mp^2}{s \,\sigma} \log 
\frac{\sigma+1}{\sigma-1} -1 \biggr) \, \log\frac{m_\gamma}{\Mp} \, \biggr\}~,
\eeqa 
and similarly for the neutral pions we find
\beqa
\frac{\Theta_1^{\pi^0} (s)}{\Theta_1^{\pi^0} (0)} &=&  1 -
\frac{1}{(4\pi F)^2} \, \biggl\{ \frac{8}{9} s +\frac{4}{3}s \log\frac{\Mp}{\Mn}
+\frac{2}{3}\Bigl(s+2\Mp^2-\Mn^2\Bigr)\, K_\pm(s) +\frac{1}{3}\Mn^2\, K_0 (s) 
\no \\
&&-\frac{4}{3}\Mn^2 \biggl(\frac{\Mp^2}{s}K_\pm (s)
+ \frac{1}{6}\biggr) + \frac{2}{3}\Mn^2 \biggl(\frac{\Mn^2}{s}K_0 (s) 
+ \frac{1}{6}\biggr) -2s \biggl( \frac{1}{3}\bar{l}_{11} +
\bar{l}_{12} \biggr) \biggr\}~,\\
\Theta_1^{\pi^0} (0) &=& 1 + \frac{\Mn^2}{(4\pi F)^2} \,
\biggl\{ \frac{1}{3} +\frac{4}{3}\log\frac{\Mp}{\Mn} +\bar{l}_{13} - 
\frac{4}{3}\bar{l}_{11} \biggl\}~,\\ 
\Theta_2^{\pi^0} (s) &=& 1 - \frac{2s}{(4\pi F)^2} \, \bar{l}_{12}~,    
\label{t120}
\eeqa 
in terms of the loop functions $K_{0,\pm} (s)$, $G(s)$ defined in 
eqs.~(\ref{lfct1}, \ref{lfct2}) and
\beqa
H(s) &=& \frac{1}{s \,\sigma} \,\biggl\{ \biggl( \frac{1}{2} \log \frac{s}{\Mp^2}
- i\pi \biggr) \, \log \frac{1+\sigma}{1-\sigma} - \int_{\tau_-}^{\tau_+} dt \,
\frac{\log t}{1-t}\,\biggr\}~,\\
H_1(s) &=& \frac{1}{\sigma^2} \,\biggl\{ \frac{1}{2} \, H(s) - \frac{1}{s}\,
\biggl( \log \frac{s}{\Mp^2} - i\pi \biggr) \biggr\}~,
\eeqa
with
\beq
\tau_\pm = \frac{1 \pm \sigma - 2\Mp^2/s}{1 \pm \sigma}~.
\eeq
The following remarks are in order. First, in the isospin limit
($m_u=m_d, \, e^2=0)$ we recover the results of ref.~\cite{DL}. 
The normalization of $\Theta_2 (0)$ is not affected by the
photon loops, in particular not by the IR divergent terms. We also
note that for the charged pions, $\Theta_1$ diverges logarithmically at $s=0$.
We have therefore only given the finite part of $\Theta_1^{\pi^\pm}
(0)$, denoted by the superscript ``fin''. This divergence is due to the
photon tadpole and self--energy graphs with coupling of the
energy--momentum tensor to the photon line. Such diagrams are not present in the
case of the scalar pion form factor simply because a scalar--isoscalar source
has no coupling to virtual photons only. We remark that after mass renormalization,
no counterterms from ${\cal L}^{(4, \, {\rm str})}$ and ${\cal L}^{(4, \,
{\rm em})}$ contribute to $\Theta_{1,2}^{\pi^a}$.
Finally, we note that the new loop functions $H(s)$ and $H_1(s)$
cancel in the matrix elements of the trace of the energy--momentum tensor as defined
in eq.~(\ref{trace}). Consequently, these matrix elements are well
behaved at $s=0$.

\medskip \noindent
We can now combine the results collected in eqs.~(\ref{t12c}--\ref{t120}) to
work out the pionic form factors of the trace of the energy--momentum tensor.
For the charged pions this leads to
\beqa
\Theta_{\pi^\pm}(s) \! &=& \! 2\Mp^2 + s \no\\
&+&\! \frac{1}{(4\pi F)^2} \,\biggl\{ \frac{2}{3}(\Mn^2-2s)s 
+ s(3s-2\Mn^2) \log\frac{\Mp}{\Mn} -\frac{1}{2}(s-\Mn^2)(s+2\Mn^2)\,K_0(s)\no\\
&&\qquad\quad - \frac{1}{2}s (s+2\Mp^2)\,K_\pm (s) + s(s-2\Mn^2)\, \bar{l}_{11}
+ 4s(s-\Mn^2)\,\bar{l}_{12} + \frac{3}{2}s\Mn^2\,\bar{l}_{13} \,\biggr\} \no\\
&+&\! \biggl(\frac{e}{4\pi}\biggr)^2 \, \biggl\{ (2\Mp^2+s)(1+4Z)\,
K_\pm (s) \no\\ && \qquad\qquad
+(2\Mp^2+s)\frac{s-2\Mp^2}{\Mp^2}\,G(s) + (3+4Z)s \, 
\biggl(\bar{k}_{15} + 2\log \frac{\Mp}{\Mn} \biggr) \no\\
&& \qquad\qquad +4\, (2\Mp^2+s) \,
\biggl( \frac{s-2\Mp^2}{s\,\sigma} \log \frac{\sigma+1}{\sigma-1} -1 \biggr)\,
\log\frac{m_\gamma}{\Mp} \, \biggr\}~,
\eeqa 
and similarly for the neutral pions
\beqa
\Theta_{\pi^0}(s) &=& 2\Mn^2 + s \no\\
&+& \frac{1}{(4\pi F)^2} \,\biggl\{ \frac{2}{3}(\Mn^2-2s)s 
-2s (s-\Mn^2) \log\frac{\Mp}{\Mn} -\frac{1}{2}\Mn^2(s+2\Mn^2)\,K_0(s)\no\\
&-& (s-\Mn^2) (s+2\Mn^2)\,K_\pm (s) + s(s-2\Mn^2)\, \bar{l}_{11}
+ 4s(s-\Mn^2)\,\bar{l}_{12} + \frac{3}{2}s\Mn^2\,\bar{l}_{13} \,\biggr\}~. \no\\ &&
\eeqa 

\medskip\noindent
One process in which these form factors play a role is the decay of a
light Higgs boson into a pion pair. The corresponding matrix element
takes the form~\cite{DGL}
\beq
\langle \pi^i (p) \pi^k (p') | {\cal L}_{\rm eff} | H\rangle =
-\frac{1}{v} \, {\cal G}(s)\, \delta^{ik}~, \quad {\cal G}(s) = 
\frac{2}{9} \Theta_\pi (s) + \frac{7}{9} \Bigl( \tilde{\Gamma}_\pi (s) + S_\pi (s) 
\Bigr)~,
\eeq
where $H$ denotes the Higgs field and the form factor $S_\pi$ is
related to the expectation value of the strangeness operator $m_s
\bar{s}s$ in the pion. In what follows, we will neglect this form
factor. Also, $v$ is the scale of electroweak symmetry breaking,
$v^{-2} = \sqrt{2}G_F \simeq (250\,$GeV)$^{-2}$ (with $G_F$ the Fermi
constant). The width of the Higgs decaying into a pair of charged
pions is then given by
\beq
\Gamma (H\to \pi^+ \pi^-) = \frac{\sqrt{2} G_F}{16 \pi m_H} \,
\sqrt{1 - \frac{4\Mp^2}{m_H^2}} \, |{\cal G} (m_H^2)|^2~,
\eeq
with $m_H$ the Higgs mass.
For more details, we refer to ref.~\cite{DGL}.

\section{Infrared regularization}
\label{sec:IR}
\def\theequation{\arabic{section}.\arabic{equation}}
\setcounter{equation}{0}

So far we have dealt with the IR divergences by introducing a
small photon mass $m_\gamma$. We are now going to cure this by
considering processes with additional soft photon radiation 
from the final--state pions with photon energies below a given 
detector resolution $\Delta E$. Before discussing the details for the various 
form factors, we have to address the power counting. In fact, for
the treatment of IR divergences, only the counting in $e$ (or,
equivalently, in the fine structure constant $\alpha$) is of
relevance. Since we consider one--photon loop diagrams, these scale as
$e^2$. Any radiation of a single soft photon carries a factor of $e$
in a given amplitude, so that, for a pertinent cross section (which is
the square of such an amplitude), we get contributions of order $e^2$.
Radiation of multiple soft photons is therefore related to higher order
photon loop diagrams. The procedure we have to follow is
standard, for a textbook treatment we refer e.g.\ to
refs.~\cite{brown,wein}. Consequently, we do not concern ourselves here
with initial state radiation.

\subsection{Vector form factor}
\begin{figure}[htb]
\centerline{
\epsfysize=6cm
\epsffile{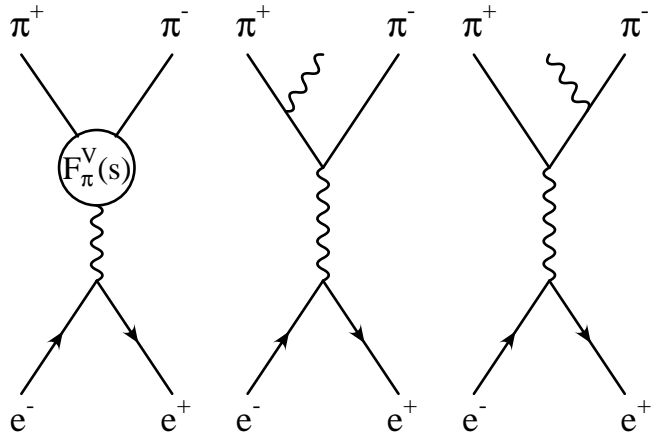}
}
\vskip 0.5cm
\caption{IR regularization of the vector form factor to one loop. 
The photon radiated from the charged pion lines is soft.
\label{Vir}
}
\end{figure}
\noindent
Consider now the vector form factor. As shown in  fig.~\ref{Vir}, it
can be measured e.g.\ in electron--positron annihilation. The
corresponding tree graphs with soft  photon radiation are also
depicted in that figure. We evaluate the square of the vector form factor.
After the IR regularization procedure, it takes  the form
\beqa
|F_\pi^V(s)|^2 &=& 
 1 + \frac{1}{(4\pi F)^2}\, \frac{1}{3} 
\biggl\{\biggl(\bar{l}_6 - \frac{1}{3}\biggr)\, s 
+  (4\Mp^2-s) \, K_\pm (s) \biggr\} \no\\ 
&&  +2 \biggl( \frac{e}{4\pi}\biggr)^2 \, 
\biggr\{\frac{1}{3} \frac{28\Mp^2-13 s}{4\Mp^2 -s} \, K_\pm (s) 
- \frac{4s}{4\Mp^2-s}
- \frac{4}{3} \biggl( \frac{\Mp^2}{s} \, K_\pm (s) +
\frac{1}{6} \biggr) \\
&&
+\frac{s-2\Mp^2}{\Mp^2} \, G(s)
+ 4 \, \biggl( \frac{s-2\Mp^2}{s\, \sigma}\,
\log \frac{\sigma+1}{\sigma-1} -1 \biggr) \, \log \frac{\Delta E}{\Mp}
+ R_V(s,\Delta E) \, \biggr\}~ \no
\eeqa
with the resolution function $ R_V(s,\Delta E)$
\beqa
R_V(s,\Delta E) &=& 
\frac{s-m_e^2}{s+2m_e^2} \frac{16}{s \,\sigma^3} \,
\int_0^{\Delta E} dl \, l \, \sqrt{ \biggl(1-\frac{2l}{\sqrt{s}}\biggr)\biggl(1
-\frac{4\Mp^2}{s}-\frac{2l}{\sqrt{s}}\biggr)} \no\\
&-& 12\frac{\Mp^2}{s+2m_e^2} \frac{1}{s \,\sigma^3} \,
\int_0^{\Delta E} dl \, l \, \sigma (l) - 4 \frac{1}{\sqrt{s}\,\sigma} \,
\int_0^{\Delta E} dl \, \log\biggl( \frac{1+\sigma (l)}{1-\sigma (l)}\biggr) \no\\
&-& \frac{s-4m_e^2}{s+2m_e^2} \, \frac{s-2\Mp^2}{s^2\,\sigma^3}\,
\int_0^{\Delta E} dl \, l\,\log\biggl( \frac{1+\sigma (l)}{1-\sigma
  (l)}\biggr ) \no\\
&-& \frac{4\Mp^2}{s\, \sigma} \, \int_0^{\Delta E} \frac{dl}{l} \,
\biggl( \frac{\sigma (l)}{1-\sigma^2 (l)} - \frac{s\,\sigma}{4\Mp^2}
\biggr)\no\\
&+& \frac{s-2\Mp^2}{s\,\sigma} \, \int_0^{\Delta E} \frac{dl}{l} \,
\log \biggl( \frac{1+\sigma (l)}{1-\sigma (l)}\,\frac{1-\sigma}{1+\sigma}
\biggr) ~,
\eeqa
where
\beq\label{sigl}
\sigma (l) = \sqrt{\frac{s-4\Mp^2-2\sqrt{s}l}{s-2\sqrt{s}l}}~,
\eeq 
$\sigma (0) = \sigma$ as defined in eq.~(\ref{sig}), and $m_e$ is the
electron mass. $\Delta E$ is the
detector resolution which should be supplied by the experimenters. Typically,
$\Delta E = 10\ldots 20\,$MeV. The resulting expression is obviously
IR finite.  For later discussion, we call the term $\sim
\log(\Delta E/\Mp)$ the IR finite contribution and the term
proportional to the resolution function  the finite
bremsstrahlung contribution. Finally, we remark that if one chooses
a different process to perform the IR regularization, the resolution
function would  be different.

\subsection{Higgs decay and scalar form factors}

\begin{figure}[htb]
\centerline{
\epsfysize=4.5cm
\epsffile{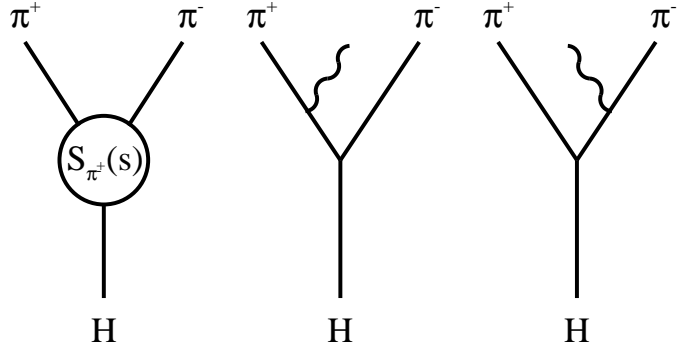}
}
\vskip 0.5cm
\caption{IR regularization of any scalar form factor $S_{\pi^\pm}(s)$ to one loop.
The solid line labeled ``H'' denotes the light Higgs particle. 
The photon radiated from the charged pion lines is soft.
\label{sir}
}
\end{figure}

We now turn to the IR regularization of the scalar form factors. Let us
genuinely denote by $S_{\pi^\pm} (s)$ any one of the charged pion scalar form factors
$\tilde{\Gamma}_{\pi^\pm} (s)$, $\Theta_{\pi^\pm} (s)$, or any linear 
combination thereof.
To the order we are working, the regularization can be performed by considering
a physical process like the decay of a light Higgs $H$ into two pions accompanied
by soft photon radiation from the final--state pions.  The isospin--conserving
part of this process has been studied in great detail in ref.~\cite{DGL}, see
also ref.~\cite{RW}. We have used that calculation as a check whenever possible. 
Of course, the calculation is somewhat academic since such a light Higgs seems to 
be ruled out by experiment. It is, however, the simplest process which allows us
to perform the IR regularization. Alternatively, one could consider processes
like $\psi' \to \psi \pi\pi$. More precisely, we
only need to consider the emission of {\it one} soft photon from any one of
the charged pions in the final--state. The method is depicted in fig.~\ref{sir}, 
which shows the tree graph for the $H\to \pi\pi$ decay via the corresponding 
scalar form factor and the two graphs with one soft photon.  Denote by 
$S^{(0)}_{\pi^\pm} (s)$ the leading term in the chiral expansion of any
scalar form factor and by $S^{(2)}_{\pi^\pm} (s)$ the 
next--to--leading order contribution.
To one--loop accuracy, the IR--divergent part of the amplitude takes the form
\beq
S^{(2,\, {\rm IR-div})}_{\pi^\pm} (s) = 4 \biggl( \frac{e}{4\pi}\biggr)^2\, 
S^{(0)}_{\pi^\pm} (s) \, \xi \, \log\frac{m_\gamma}{\Mp}~,
\eeq
with
\beq
\xi = \frac{s-2\Mp^2}{s\,\sigma} \,\log\frac{\sigma+1}{\sigma-1} -1~.
\eeq
With that, we can write the square of the form factor as
\beqa\label{Sreg}
\bigl|S_{\pi^\pm} (s)\bigr|^2 &=&
\bigl|S^{(0)}_{\pi^\pm} (s)\bigr|^2 \, \biggl\{ 1 + 8\biggl( \frac{e}{4\pi}\biggr)^2 
\,\xi\,
\log\frac{\Delta E}{\Mp} + \biggl( \frac{e}{4\pi}\biggr)^2
\, R_S(s,\Delta E) \,\biggr\}\no\\
&& 
+2 \,{\rm Re}\, S^{(0)}_{\pi^\pm} (s) \, 
S^{(2, \, {\rm fin})}_{\pi^\pm} (s) + {\cal O}(e^4)~,
\eeqa
with the resolution function $R_S(s,\Delta E)$ given by
\beqa
R_S(s,\Delta E) &=& -\frac{16}{s \, \sigma} \, \biggl\{ 2\Mp^2 \,
\int_0^{\Delta E} \frac{dl}{l} \, \biggl( \frac{\sigma (l)}{1-\sigma^2 (l)}
- \frac{s\,\sigma}{4\Mp^2} \biggr) +\sqrt{s}\, \int_0^{\Delta E} {dl}\, \log
\biggl( \frac{1+\sigma (l)}{1-\sigma (l)} \biggr) \no\\
&& \qquad \quad+ \biggl(\Mp^2 - \frac{s}{2}\biggr) \, 
\int_0^{\Delta E} \frac{dl}{l} \,
\log \biggl( \frac{1+\sigma (l)}{1-\sigma (l)}\, \frac{1-\sigma}{1+\sigma}\,
 \biggr) \biggr\}~,
\eeqa
where $\sigma (l)$ has been defined in eq.~(\ref{sigl})
and $\sigma (0) = \sigma$ in eq.~(\ref{sig}), respectively. 
Again, the term $\sim \xi$ leads to the
IR finite pieces and the term $\sim R_S$ to the finite bremsstrahlung
contribution. We refrain from writing down the explicit IR finite
expressions for the various scalar form factors since they can be deduced easily 
from eq.~(\ref{Sreg}) and the formulae spelled out in
sections~\ref{sec:pion}~and~\ref{sec:EMT}.

\section{Results and discussion}
\label{sec:res}
\def\theequation{\arabic{section}.\arabic{equation}}
\setcounter{equation}{0}
\subsection{Vector form factor and pion charge radius}
\begin{figure}[hbt]
\centerline{
\epsfysize=8cm
\epsffile{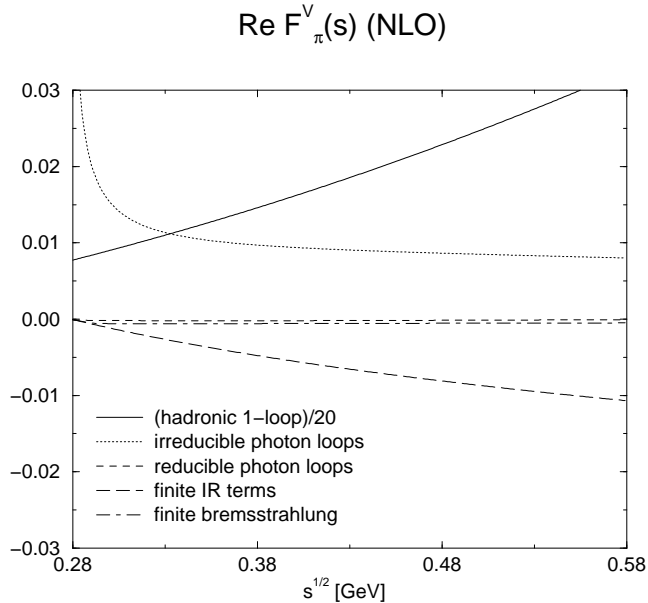}
}
\vskip 0.2cm
\caption{Vector form factor in the time--like region. The various
virtual photon contributions are shown in comparison with the strong
fourth order contribution. The latter is divided by a factor of 20.
\label{FVs}
}
\end{figure}
\begin{figure}[htb]
\centerline{
\epsfysize=8cm
\epsffile{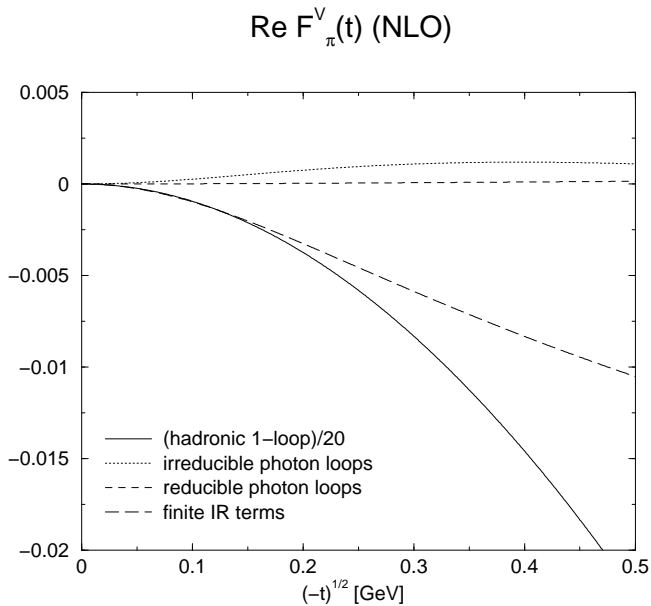}
}
\vskip 0.2cm
\caption{Vector form factor in the space--like region. The various
virtual photon contributions are shown in comparison with the strong
fourth order contribution. The latter is divided by a factor of 20.
\label{FVt}
}
\end{figure}

\noindent
In fig.~\ref{FVs}, we show the vector form factor in the time--like
region. We remark that the electromagnetic corrections from the
reducible photon loops and from the bremsstrahlung are very small, the
latter contribution being so tiny that it  will be
discarded in what follows. A bit above threshold, 
the finite IR terms are of similar size as
the irreducible photon loop contribution. For comparison, we have also
shown the strong fourth order contribution divided by a factor of
20. We conclude that the electromagnetic corrections are of the expected
size of about one percent, except for the divergence of the photon loop graphs at
$s=4\Mp^2$. In the space--like region, shown in fig.~\ref{FVt}, the
electromagnetic effects from the finite IR graphs are somewhat more
pronounced, the photon loop contribution is small. From this,
we can deduce the electromagnetic corrections to the pion charge radius,
\beqa
\langle r^2 \rangle_{\pi, {\rm strong}}^V &=& \frac{1}{(4\pi F)^2} \,
\biggl( \bar{l}_6 -1 \biggr)~,\\
\langle r^2 \rangle_{\pi, {\rm photon}}^V &=& - \biggl( \frac{e}{4\pi}\biggr)^2
\, \frac{31}{5\Mp^2}~,\\\label{rIR}
\langle r^2 \rangle_{\pi, {\rm finite~IR}}^V &=& -\biggl( \frac{e}{4\pi}\biggr)^2 \,
\frac{8}{\Mp^2}\, \log \frac{\Delta E}{\Mp}~.
\eeqa
The numerical results for these various contributions are
\beq
\langle r^2 \rangle_{\pi, {\rm strong}}^V  = 0.442~{\rm fm}^2\,,\,\,
\langle r^2 \rangle_{\pi, {\rm photon}}^V = -0.007~{\rm fm}^2\,,\,\,
\langle r^2 \rangle_{\pi, {\rm finite~IR}}^V = 0.024\, (0.018)~{\rm fm}^2~,
\eeq
for $\Delta E =10 \, (20)$~MeV. We use $F = 93\,$MeV, $\Mp =139.57\,$MeV 
and $\bar{l}_6 = 16.5$. There are two ways of dealing with the
resolution dependent IR contribution. First, if one compares the
similar process $e^+ e^- \to \mu^+ \mu^-$, soft photon radiation to
leading order in $e^2$ gives a finite IR contribution
to the ``intrinsic'' muon radius, which is believed to be zero. This
contribution has the same form as in eq.~(\ref{rIR}) but with the
pion mass replaced by the muon mass (see e.g.\ the explicit calculation
in ref.~\cite{brown}). If one therefore directly
compares the processes $e^+ e^- \to \pi^+ \pi^- (\gamma)$ and $e^+ e^- \to \mu^+
\mu^- (\gamma)$, one is only sensitive to the difference $(1/\Mp)\log(\Delta
E/\Mp) - (1/M_\mu)\log(\Delta E/M_\mu)$ which reduces the dependence
on the detector resolution by almost one order of magnitude. Another
way of proceeding is to follow the arguments which have been used in
the discussion of $\pi\pi$ scattering~\cite{RS,KU}. While the cross
section of course depends on the detector resolution and so does the
scattering length $a$, $d\sigma / d\Omega = |a|^2$, one can define
an electromagnetically corrected scattering length which is
independent of $\Delta E$. This is achieved by absorbing the
resolution dependent terms into the appropriately redefined phase
space. Such a procedure applied here would essentially amount to
dropping the term given in eq.~(\ref{rIR}).
To summarise, the resolution dependent terms are probably not to be 
understood as ``real'' isospin violation effects in the reaction
$e^+ e^- \to \pi^+ \pi^-$, but rather as part of the radiative process 
$e^+ e^- \to \pi^+ \pi^- \gamma$, and therefore should not be regarded
as intrinsic electromagnetic corrections to the form factor.

\medskip \noindent
We should explicitly compare these numbers to the two--loop corrections
for the vector form factor calculated in ref.~\cite{BT}. Depending on the set
of parameters they use, they find corrections at ${\cal O}(q^6)$ which 
numerically amount to
\beq
\langle r^2 \rangle^V_{\pi,{\rm 2-loop}} = 0.017 \ldots 0.022~{\rm fm}^2~.
\eeq
This is larger than the ``intrinsic'' virtual photon contribution
found above by about a factor of 3, and
thus is similar to the findings in the pion--pion scattering
case, see refs.~\cite{MMS,KU}.

\medskip \noindent
In refs.~\cite{GM,BT}, also the quadratic term in the Taylor 
expansion of $F_\pi^V(s)$ according to 
\beq
F_\pi^V(s) = 1 + \frac{1}{6}\langle r^2 \rangle_\pi^V s + c_\pi^V s^2
  +{\cal O}(s^3)
\eeq
was considered.
Counterterms to fit experimental values of $c_\pi^V$ exactly
only appear at ${\cal O}(q^6)$, and indeed the finding in \cite{GM,BT} is that
the ${\cal O}(q^4)$ calculation is completely insufficient to describe $c_\pi^V$, with
the final value dominated by an ${\cal O}(q^6)$ counterterm. 
Here, we might either check whether this large ${\cal O}(q^6)$ contribution can be reduced
when taking into account the electromagnetic corrections, or  we can
give credit to the two--loop analysis in case these
again turn out to be very small. The contributions found from expanding
eq.~(\ref{FV1loop}) are
\beqa
c^V_{\pi,{\rm strong}} &=& \frac{1}{(4\pi F)^2} \,\frac{1}{60 \Mp^2}
= 0.626~{\rm GeV}^{-4},\\
c^V_{\pi,{\rm photon}} &=& - \biggl(\frac{e}{4\pi}\biggr)^2 \frac{113}{420 \Mp^4}
= -0.412~{\rm GeV}^{-4},\\
c^V_{\pi,{\rm finite IR}} &=& - \biggl(\frac{e}{4\pi}\biggr)^2 \frac{1}{5 \Mp^4}
\log \frac{\Delta E}{\Mp} = 0.807\,(0.595)~{\rm GeV}^{-4},
\eeqa
for $\Delta E =10 \, (20)$~MeV, 
whereas, for comparison, the authors of \cite{BT} give a value of
\beq
c^V_\pi = (3.85 \pm 0.60)~{\rm GeV}^{-4}~,
\eeq
found by a fit of the ${\cal O}(q^6)$ CHPT amplitude to the data. We see
therefore that the electromagnetic effects on the curvature 
are of a size comparable to the hadronic one--loop parts, although
resolution dependent and independent contributions tend to cancel to some extent
for typical values of $\Delta E$.
In the analysis of ref.~\cite{BT}, this would lead to a moderate change in their
LEC $r^r_{V2}$.

\medskip \noindent 
We conclude that, throughout, the 
intrinsic virtual photon effects on the pion vector form factor
in the low energy region are of the expected size and smaller than the two
loop effects calculated in refs.~\cite{GM,BT}. (Note finally that one
particular $SU(3)$ effect (the charged to neutral kaon mass difference
in kaon loops) has also been found to be tiny, see ref.~\cite{aus}.)
 
\subsection{Scalar form factor and scalar pion radius}
\noindent
We now turn to the scalar form factor. In contrast to the vector form
factor, we have an additional electromagnetic effect, namely the
pion mass difference in the hadronic loops (of course, there is also
a tiny strong contribution to the pion mass splitting). 
\begin{figure}[htb]
\centerline{
\epsfysize=8cm
\epsffile{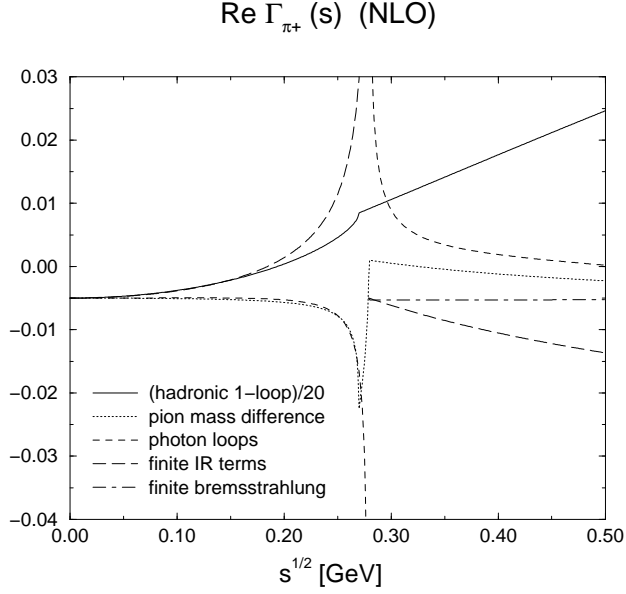}
}
\vskip 0.2cm
\caption{Scalar form factor of the charged pions in the time--like
  region. The various
virtual photon contributions are shown in comparison with the strong
fourth order contribution. The latter is divided by a factor of 20.
\label{FSc}
}
\end{figure}

\medskip \noindent
Consider the charged pions first. The various virtual photon
corrections are shown in fig.~\ref{FSc} in comparison with the strong
fourth order contribution scaled down by a factor of 20. Here,
the different contributions are somewhat more pronounced as compared
to the vector case, but again we observe cancellations between the
various effects. Note that the finite bremsstrahlung contribution is
again very small. The corresponding corrections to the scalar radius
of the pion can be worked out easily (we again neglect the tiny
bremsstrahlung piece),
\beqa
\langle r^2 \rangle_{\pi^\pm, {\rm strong}}^S &=& \frac{6}{(4\pi F)^2} \,
\biggl(\bar{l}_4 - \frac{13}{12} \biggr)~,\\
\langle r^2 \rangle_{\pi^\pm, {\rm mass}}^S &=& -\frac{6}{(4\pi F)^2} \,
\log\frac{\Mp}{\Mn} + \biggl( \frac{e}{4\pi}\biggr)^2 \frac{4 Z}{\Mp^2}~,\\
\langle r^2 \rangle_{\pi^\pm, {\rm photon}}^S &=& \biggl( \frac{e}{4\pi}\biggr)^2
\, \frac{1}{\Mp^2}~,\\
\langle r^2 \rangle_{\pi^\pm, {\rm finite~IR}}^S &=& - \biggl( \frac{e}{4\pi}\biggr)^2 
\, \frac{8}{\Mp^2}\, \log \frac{\Delta E}{\Mp} ~.\label{rIRS}
\eeqa
The novel contribution due to the pion mass difference in the pion loops
is denoted by the subscript ``mass''.
The resolution dependent term is identical to the one
obtained for the vector form factor. It can be dealt with in the
same manner as described in the preceding paragraph. The
corresponding values are
\beqa
\langle r^2 \rangle_{\pi^\pm, {\rm strong}}^S  &=& 0.550~{\rm fm}^2\,,
\quad \,\,\,
\langle r^2 \rangle_{\pi^\pm, {\rm mass}}^S = -0.0016~{\rm fm}^2\,, \no\\ 
\langle r^2 \rangle_{\pi^\pm, {\rm photon}}^S &=& 0.0012~{\rm fm}^2\,, \quad
\langle r^2 \rangle_{\pi^\pm, {\rm finite~IR}}^S = 0.024\, (0.018)~{\rm
  fm}^2~,
\eeqa
for $\Delta E =10 \, (20)$~MeV and using $\bar{l}_4 =4.3$.
We note that the small corrections due
to the pion mass difference ($-0.3\%$) and the photon loops ($+0.2\%$)
nearly cancel each other to give a tiny effect on the radius. 
Even the separate contributions are, again, by one order of magnitude smaller
than the two--loop corrections \cite{BT},
\beq
\langle r^2 \rangle^S_{\pi,{\rm 2-loop}} = 0.016 \ldots 0.025~{\rm fm}^2~.
\eeq
The normalization of the scalar form factor is also of interest. We 
normalize it here to the mass of the charged pions (although the canonical
reference mass is the one of the neutral pion),
\beq
\tilde{\Gamma}_{\pi^\pm} (0) = \Mp^2 \, ( 1 - 0.065 - 0.012 - 0.004)~,
\eeq
where the first correction is due to the electromagnetic operator $\sim C$, i.e.
the pion mass difference, the second term is the well--known hadronic
shift~\cite{GLAnn} (using $\bar{l}_3 =2.9$), and the third term comprises the genuine
electromagnetic corrections.
To arrive at this last number, we have used the dimensional analysis of ref.~\cite{KU},
$|k_i^r (M_\rho)| \leq 1/(4\pi)^2$, which leads to $|\bar{k}_\pm '| \leq 6.3$.
Consequently, the shift due to the electromagnetic corrections is about one third
of the strong correction to the normalization of the charged pions scalar form
factor.

\medskip \noindent
As in the case of the vector form factor, we wish to give credit to the analysis
of the quadratic term of the Taylor expansion of the form factors given in 
refs.~\cite{GM,BT}. The remarks made about the purpose of such an analysis for
the one--loop electromagnetic corrections in the case of $F_\pi^V(s)$ apply
equally here. From eq.~(\ref{GSch1loop}) we find
\beqa
c^S_{\pi^\pm,{\rm strong}} &=& \frac{1}{(4\pi F)^2}\,\frac{19}{120\Mn^2}
= 6.36~{\rm GeV}^{-4}~, \\
c^S_{\pi^\pm,{\rm mass}} &=& \biggl(\frac{e}{4\pi}\biggr)^2 \frac{Z}{\Mp^2}
\biggl( \frac{1}{15\Mp^2} - \frac{1}{6\Mn^2} \biggr)
= -0.15~{\rm GeV}^{-4}~, \\
c^S_{\pi^\pm,{\rm photon}} &=& - \biggl(\frac{e}{4\pi}\biggr)^2
\frac{1}{12\Mp^4} = -0.13~{\rm GeV}^{-4}~, \\
c^S_{\pi^\pm,{\rm finite IR}} &=& - \biggl(\frac{e}{4\pi}\biggr)^2
\frac{1}{5\Mp^4} \log \frac{\Delta E}{\Mp} = 0.81\,(0.60)~{\rm GeV}^{-4}~.
\eeqa
Note that the difference in the numerical value for the hadronic one--loop
contribution as compared to ref.~\cite{BT} occurs because they employ the charged
pion mass whereas, when taking isospin violation into account, the correct
reference mass is the neutral pion mass. The value extracted from experimental
data is \cite{GM,BT}
\beq
c^S_\pi = (10.6 \pm 0.6)~{\rm GeV}^{-4}~,
\eeq
with the error taken from the range of different analyses. Again, there have to
be substantial ${\cal O}(q^6)$ contributions, the electromagnetic corrections
are small and will lead only to small modifications in the analysis of the 
${\cal O}(q^6)$ counterterms.

\begin{figure}[htb]
\centerline{
\epsfysize=8cm
\epsffile{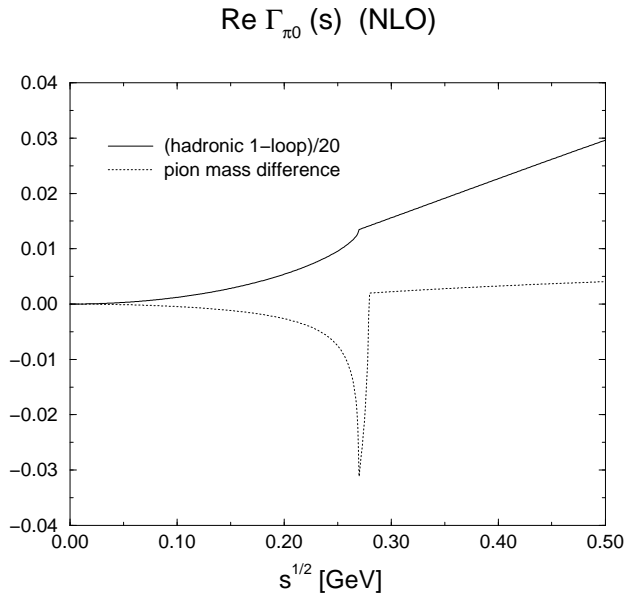}
}
\vskip 0.2cm
\caption{Scalar form factor of the neutral pions in the time--like
  region. The only electromagnetic contribution is due to the pion
  mass difference. It is shown in comparison with the strong
fourth order contribution. The latter is divided by a factor of 20.
\label{FSn}
}
\end{figure}

\medskip \noindent
We now turn to the neutral pions. Here, the only effect is due to
the pion mass difference in the pion loops, see
fig.~\ref{FSn}. The corresponding contribution to the radius takes the form
\beq
\langle r^2 \rangle_{\pi^0, {\rm mass}}^S = -\frac{12}{(4\pi F)^2} \,
\log\frac{\Mp}{\Mn} + \biggl( \frac{e}{4\pi}\biggr)^2 \frac{2 Z}{\Mp^2}~.
\eeq
Numerically we have
\beq
\langle r^2 \rangle_{\pi^0, {\rm mass}}^S = -0.0094~{\rm fm}^2\,,
\eeq
using $\Mn = 134.97\,$MeV. This amounts to a 2\% reduction of the
hadronic result. Such an  enhancement
of the electromagnetic effects for the neutral particles as compared
to the charged ones has also been observed in pion--nucleon
scattering~\cite{W77,MS,FMS,MM} or neutral pion photoproduction off 
nucleons~\cite{bkmpi0}. Finally, we study the normalization.
We find
\beq
\tilde{\Gamma}_{\pi^0} (0) = \Mn^2 \, ( 1 - 0.013 - 0.004 - 0.001)~,
\eeq
where the first correction is due to the hadronic shift~\cite{GLAnn},
the second one is  due to the strong isospin breaking $\sim \bar{l}_7$
(we use $\bar{l}_7 =1.0$), and the third term stems from the pion mass 
difference in the loops.  We
note that the strong isospin breaking  amounts to a 30\% correction
of the strong isospin conserving shift.  

\medskip \noindent
To conclude this section, let us quote the result for the isospin--breaking
one--loop contribution to the parameter $c^S_{\pi^0}$,
\beq
c^S_{\pi^0,{\rm mass}} = \frac{1}{(4\pi F)^2}\frac{1}{\Mp^2}\biggl(\frac{1}{6}
-\frac{3}{20}\frac{\Mp^2}{\Mn^2}-\frac{1}{60}\frac{\Mn^2}{\Mp^2}\biggr)
= -0.35~{\rm GeV}^{-4}~,
\eeq
which is again slightly larger than for the charged pions, but small compared
to the two--loop contribution.

\subsection{Form factors of the energy--momentum tensor}

\begin{figure}[htb]
\centerline{
\epsfysize=8cm
\epsffile{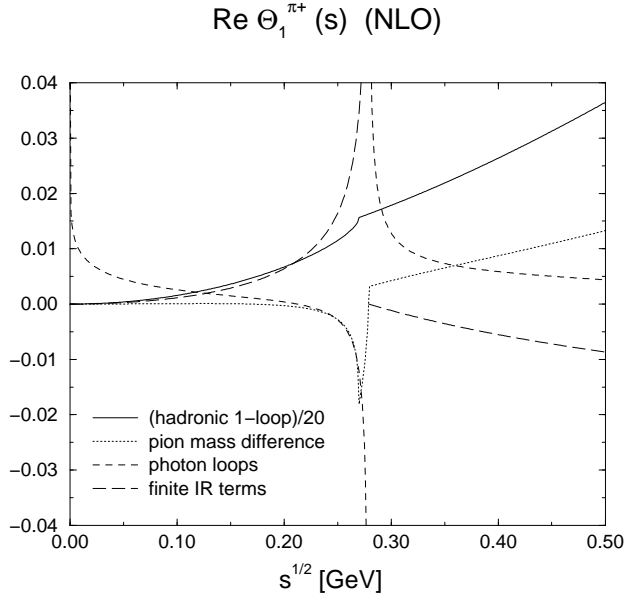}
}
\vskip 0.2cm
\caption{Normalized form factor $\Theta_1^{\pi^\pm} (s)$ of the energy--momentum tensor
for charged  pions. The various electromagnetic corrections
are shown in comparison with the strong fourth order contribution. 
The latter is divided by a factor of 20.
\label{T1c}
}
\end{figure}
\begin{figure}[htb]
\centerline{
\epsfysize=8cm
\epsffile{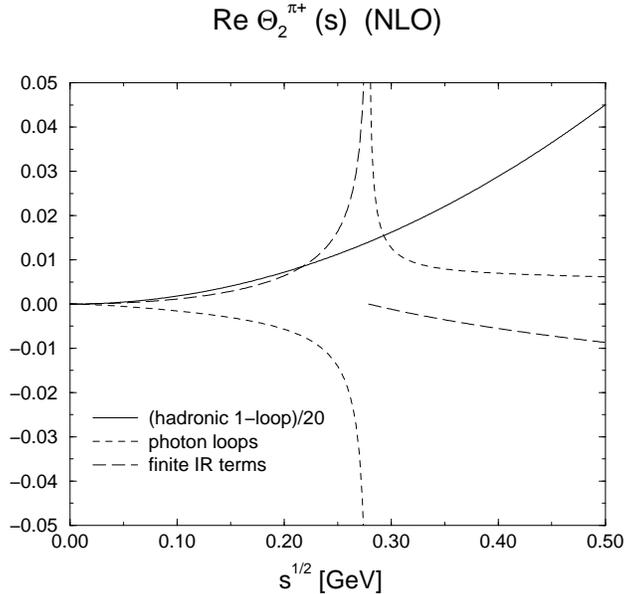}
}
\vskip 0.2cm
\caption{Tensor form factor of the energy--momentum tensor
for charged  pions. The various electromagnetic corrections
are shown in comparison with the strong fourth order contribution. 
The latter is divided by a factor of 20.
\label{T2c}
}
\end{figure}

\noindent
We have calculated  the various form factors of the energy--momentum
tensor for charged and neutral pions discussed in sect.~\ref{sec:EMT}.
We refrain from discussing all of them in detail but rather show two typical
cases. $\Theta_1^{\pi^\pm} (s)$ is given in fig.~\ref{T1c}. The electromagnetic
corrections look similar to the ones for the scalar form factor shown in
fig.~\ref{FSc} with two notable exceptions. First, as noted before, the 
photon loop corrections diverge at $s=0$, and second, the corrections due to
the pion mass difference in the pion loops grow with increasing $s$ after the 
two--pion threshold. The tensor form factor for charged pions is shown in
fig.~\ref{T2c}. In this case, the hadronic fourth order correction is simply
a polynomial linear in $s$ since there are no loop contributions. The size of the
electromagnetic corrections is typically a few percent of the strong
fourth order contribution, with the exception of the kinematical points where
divergences appear. Finally, we note that the form factors of the trace of
the energy--momentum tensor are very similar to the scalar form factors 
discussed in the previous paragraph.

\section{Summary}
\label{sec:sum}
\def\theequation{\arabic{section}.\arabic{equation}}
\setcounter{equation}{0}

In this paper, we have considered various pion form factors in the presence
of virtual photons at next--to--leading order in $SU(2)$ chiral perturbation
theory. The pertinent results of this investigation can be summarized as follows:
\begin{enumerate}
\item[(1)] The vector form factor of the pion has been calculated including 
photon loops and electromagnetic counterterms. The  electromagnetic corrections
are typically a few percent of the strong fourth order contribution, 
cf.\ figs.~\ref{FVs}, \ref{FVt}. 
The intrinsic virtual photon corrections to the pion charge radius 
is of the order of 1\%.
In addition, there is a detector resolution dependent contribution to the
radius which is of the order of 5\% for $\Delta E = 10\ldots 20\,$MeV. We have
discussed how to deal with this contribution.
\item[(2)] We have also investigated the scalar form factor of the charged and the
neutral pions. For the charged pions, the results are similar to the vector form
factor, cf.\ fig.~\ref{FSc}. For neutral pions, the only effect is due to the 
pion mass difference in the pion loops. This leads to a 2\% reduction of the 
scalar radius of the neutral pion. We have also found that the scalar form 
factors at zero momentum transfer are more strongly affected by isospin breaking.
\item[(3)] The pion matrix elements of the energy--momentum tensor are parametrized
by a scalar and a tensor form factor. We have extended the chiral Lagrangian coupled
to gravity in the presence of virtual photons. There is only one additional
electromagnetic operator at fourth order. Since the energy--momentum tensor
can couple directly to virtual photons, one of the corresponding form factors
shows a divergence at zero momentum transfer not present in the scalar form
factor of the pion. Typically, electromagnetic corrections are small, see
figs.~\ref{T1c}, \ref{T2c}.
\end{enumerate}
We conclude by noting that presently the largest uncertainty in the
low energy representation of the pion form factors is due to the 
uncertainties in the low energy constants $\bar{l}_4$ and
$\bar{l}_6$. Strong two--loop corrections or the virtual photon
effects calculated here are sizeably smaller.

\subsection*{Acknowledgements}
\noindent
We are grateful to K. Melnikov for pointing out an error in the 
previous version.

\vspace{3cm}


\end{document}